\documentclass[12pt]{article}
\usepackage{latexsym,amsmath,amssymb,amsbsy,graphicx}
\usepackage[cp1251]{inputenc}
\usepackage[T2A]{fontenc}
\usepackage[space]{cite} 
\usepackage[centerlast,small]{caption2}%
\makeatletter\def\@biblabel#1{\hfill#1.}\makeatother

\allowdisplaybreaks \multlinegap0pt 
\begin {document}
{\flushleft } \noindent\begin{minipage}{\textwidth}
{\small Astrophysics {\bf 63,} 91--107 (2020)\,\,\,\,\,\,\,\,\,\,\,\,\,\,\,\,\,\,\,\,\,\,\,\,\,\,\,\,\,\,\,\,\,\,\,\,\,\,\,\,\,\,\,\,\,\,\,\,\,\,\,\,\,\,\,\,\,\,\,\,\,\,\,\,\,\,\,\,\,\,\,\,\,\,\,\,\,\,\,\,\,\,\,\,\,\,\,\,\,\,\,\, doi:\,10.1007/s10511-020-09617-4 \\
{\scriptsize{\copyright\,2020. Springer Science+Business Media, LLC.
Translated from Astrofizika (February 2020).}}}
\begin{center}
\bigskip
{\normalsize{\bfseries ON THE ORIGIN OF OPTICAL RADIATION DURING THE
IMPULSIVE PHASE
OF FLARES ON dMe STARS.\\
I. DISCUSSION OF GAS DYNAMIC MODELS}}\\[7pt]
\end{center}
{\par \hspace*{1.2cm}\bfseries E. S. Morchenko} {\par
\hspace*{1.2cm}\footnotesize{Physics Faculty, M. V. Lomonosov Moscow
State University, Moscow, Russia;\\
\hspace*{1.1cm} e-mail: \texttt{morchenko@physics.msu.ru}}}
{\par \hspace*{1.2cm}\scriptsize Original article submitted August 8, 2019; accepted for publication December 18, 2019}\\

\setlength{\leftskip}{1.2cm} { \footnotesize{\it In connection with
a published critique, the author justifies the use of a motionless
homogeneous plane layer of pure hydrogen plasma that is near local
thermodynamic equilibrium (LTE) for analyzing the characteristics of
the radiation from a chromospheric condensation of thickness
$\Delta{}z_m=10\text{ km}$ in a gas dynamic model of stellar flares.
It is shown that the shock-wave model of flares proposed by Belova
and Bychkov, as opposed to the model of Kostyuk and Pikel’ner, has
irremovable internal defects owing to exclusion of the interaction
between a thermal wave (temperature jump) and a non-stationary
radiative shock. In particular, this model (a) does not make it
possible to increase the geometric thickness of a chromospheric
condensation owing to divergence of the fronts of the thermal and
shock waves during impulsive heating, (b) cannot provide heating of
the chromospheres of  red dwarfs over significant distances, and (с)
predicts ${H}_{\alpha}$ line profiles in conflict with observational
data. It is argued that: (a) the shock-wave model by Belova and
Bychkov represents a development of the kinematic model of solar
flares (Nakagawa et al.) and its application to dMe stars,
specifically: a study of the radiative response of the chromosphere
of a red dwarf to impulsive heating in the simplest gas dynamic
statement of the problem (a thermal wave is excluded, a stationary
approach is used); (b) in terms of the Kostyuk and Pikel’ner model,
the regions behind the stationary shock fronts do not correspond to
a chromospheric condensation with time-varying thickness but to
zones in which the plasma relaxes to a state of thermal equilibrium.
It is emphasized that the separation of the Kostyuk and Pikel'ner
model into ``thermal'' and ``shock-wave'' components is
fundamentally impossible.

}}

{\footnotesize{ Keywords: {\it red dwarf stars: flares: impulsive
heating: gas dynamic models: optical radiation}\vspace{1pt}\par}}
\bigskip
\bigskip
\end{minipage}\\
{\small{\bfseries{1. Introduction.}} In a recent article
 Belova and Bychkov \cite{Bychkov19}  calculated the profiles of
plane-parallel stationary radiative shock waves propagating in the
chromosphere of a red dwarf in the direction of the photosphere
(``downward'') at a velocity of 30--100 km/s (referred to as the
shock-wave model of stellar flares below). They took into account
the difference in the heating of the atom-ion and electron
components of the plasma behind a stationary shock front
\cite{Pik54,Mor15} ($T_{ai}>T_e$, where $T_{ai}$  is the atom-ion
temperature and $T_e$ is the electron temperature). Their
calculations \cite{Bychkov19} neglect the influx of energy from
non-thermal electrons (heating power $P_e=0$)\footnote{A constant
energy input is a necessary condition for ``downward'' propagation
of a shock wave in the chromosphere of the Sun and dMe stars over a
long time (see Ref. 4).},
 thermal conductivity (classical thermal flux $F_c=0$),
 gravitational acceleration
($g=0$), and the time-independent power of the heating sources that
maintain the stationary state of the unperturbed chromosphere
($Q=0$). The effect of the hot post-shock plasma on the degree of
ionization of the cold gas ahead of the front (precursor) is
neglected; until the passage of the shock wave the plasma is
uniform. The calculations \cite{Bychkov19} were performed in a
reference system associated with the discontinuity (viscous jump).
The magnetic field ($H_0$ from 0 to 5 G \cite{Bychkov19}) is
directed perpendicular to the velocity ($u_0$) of the unperturbed
gas; the frozen-in condition holds not only at the viscous jump but
also in the flow over the entire duration of non-stationary plasma
cooling.

Based on these calculations the authors \cite{Bychkov19} have shown
that under the conditions of the chromospheres of dMe stars ``the
gas behind the front [of a stationary shock wave] remains
transparent in the optical continuum…'' \cite{Bychkov19} and it was
concluded that during a flare the ``emission in [hydrogen] lines is
determined by a shock wave in layers above the photosphere, while
black-body radiation comes from the photosphere that is heated by a
flux of suprathermal particles'' \cite{Bychkov19}. Belova and
Bychkov also assume \cite{Bychkov19} that ``the model of a shock
wave propagating in the chromospheric gas may be applicable to
explaining the radiation from a [flaring] hydrogen plasma that is
transparent in frequencies of the continuum spectrum.''

Belova and Bychkov have also criticized (see
\cite{Bychkov17,Bychkov18} and \cite{Bychkov19}) an article by
Katsova {\it et al.} \cite{Katsova} that discusses the results of a
study of the response of a red-dwarf chromosphere to impulsive
heating by a beam of accelerated electrons with a power-law spectrum
(low-energy cutoff at $E_{10}=10{\text{ keV}}$, spectral index
$\gamma=3$ (hard spectrum), energy flux at the upper boundary of the
flare region of $F_0=10^{12}\text{ erg/cm}^2{\text{s}}$, heating
duration 10 s, and a rectangular impulse). Thus, in \cite{Bychkov17}
the authors \cite{Bychkov19} pointed out that Katsova {\it et al.}
\cite{Katsova} are using  a ``quasi-stationary approximation'' for
calculating the atomic level populations ($n_k$, where $k$ is the
principal quantum number) in terms of which the values of $n_k$ are
uniquely specified by instantaneous value of the temperature
 $T_{ai}=T_e=T$ (see Eq. (\ref{Eq1}) of this paper), while ``for emission behind the front of a [stationary] shock under the conditions
 of the chromospheres of stars in late spectral classes, the populations of the discrete
levels of a hydrogen atom are determined by instantaneous
temperature and electron density $n_e$, but also depend on the
entire prehistory of the process, beginning with heating at the
shock front'' \cite{Bychkov17}. Belova and Bychkov pointed out
\cite{Bychkov19} that ``for computing the absorption coefficient the
authors \cite{Katsova} use calculations
 … that are valid for stellar atmospheres under of thermodynamic
 equilibrium,'' ``while the situation behind the shock front [propagating at a constant velocity] is not only out of equilibrium, but also
 non-stationary.'' Finally, in \cite{Bychkov18} they \cite{Bychkov19} noted that Katsova {\it et al.}
 \cite{Katsova} use a model of the hydrogen atom consisting of just
two levels (+ continuum).

Based on solving a system of balance equations for elementary
processes \cite{Mor15} it has been shown \cite{Mor16} that the
Menzel factors for the atomic levels of a gas in a motionless
homogeneous plane layer with $T_{ai}=T_e$ corresponding
\cite{Katsova} to a chromospheric condensation\footnote{The dense
cold formation between the thermal wave front (temperature jump) and
the relaxation zone of the plasma to a state of thermal equilibrium
behind the front of a non-stationary shock wave (see Fig. 1).}
\begin{centering}
\begin{figure}\label{Fig1} 
\includegraphics[scale=1.25]{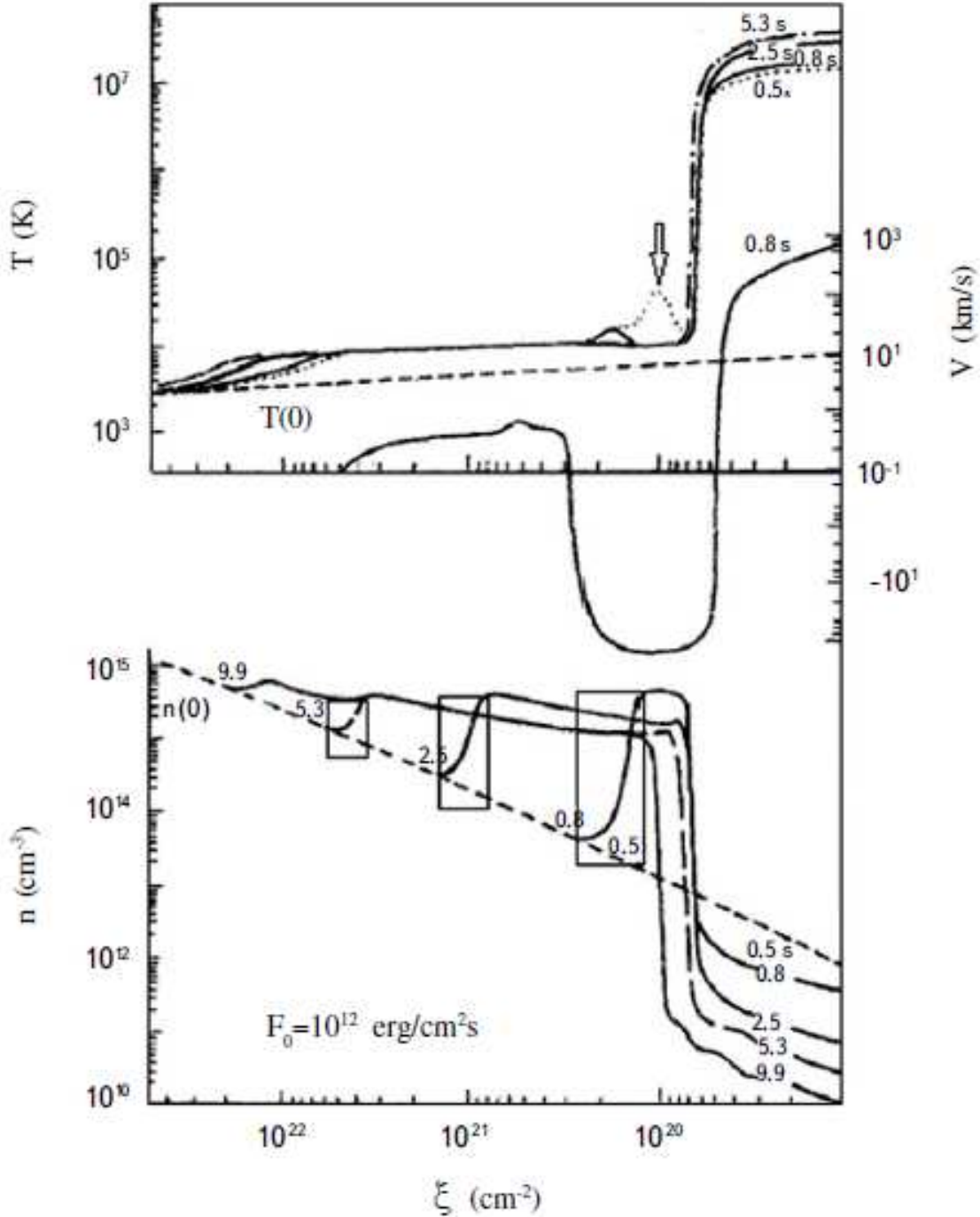}
\caption{\footnotesize ``Profiles of the density, temperature, and
velocity [of the gas] at different times'' (Katsova {\it et al.}
\cite{Katsova}). Here $n\equiv{n_\mathrm{H}}$ is the total density
of hydrogen atoms and protons and $\xi$ is the Lagrangian
coordinate: $d\xi=-n_{\mathrm{H}} dz$ \cite{Katsova}, where $z$ is
the height above the photosphere. The concentration $10^{15}$ should
be read as $10^{16}$. The rectangles distinguish the range of values
of $\xi$ corresponding to the relaxation region of the plasma at
three points of time. The arrow denotes heating of the gas directly
behind the shock front ($T_{ai}=T_e$ \cite{Kos86}). The smooth
temperature profile on the right and the narrow transition zone
after it follow from including the term $F_c$
($\varkappa_e\propto{}T_e^{5/2}$ \cite{Somov80} and $\varkappa_e$ is
the electron thermal conductivity). ``Positive values of the
velocity correspond to removal of plasma from the star's surface.
The photosphere lies to the left… The dashed curve is the initial
model [of the atmosphere]'' \cite{Katsova}.}
\end{figure}
\end{centering}
of thickness $\Delta{}z_m=10\text{ km}$ differ little from unity,
and the emission from such layer is transparent in the optical
continuum. This fact was viewed \cite{Mor16} as an important
argument in favor of the viewpoint of Grinin and Sobolev
\cite{Grin77} on the formation of the quasi-black-body radiation
observed at the brightness maximum of powerful stellar flares (the
blue component of the optical continuum) near the
photosphere.\footnote{The term ``black body'' is applied in
\cite{Mor16} to the blue component of the optical continuum of
powerful flares. In reality, this component of the radiation is
quasi-black-body \cite{Grin77}.} Thus, the critical comments of
Belova and Bychkov (\cite{Bychkov17,Bychkov18} and \cite{Bychkov19})
regarding the work of Katsova {\it et al.} \cite{Katsova} apply
substantially to the article by Morchenko \cite{Mor16}, as well.
Also the astrophysical conclusion \cite{Bychkov19} about the source
location of the blue continuum of flares in stellar photospheres
which literally regenerated \cite{Bychkov19} the concept of Gordon
and Kron \cite{Kron}, does not agree with the point of view of Refs.
11 and 8.

In this same article \cite{Mor16} it was noted that the heating of
the atom-ion component (of the gas) along the Hugoniot adiabatic and
the electron component along the Poisson adiabatic
\cite{Pik54,Mor15} is not only inherent to stationary shocks, but
also to non-stationary ones. As a result, immediately behind the
shock front \cite{Katsova} in the region referred to in \cite{Kos86}
as the ``relaxation zone [of the plasma] to a state of thermal
equilibrium,'' $T_{ai}\gg{T_e}$ from the beginning (this situation
was first noted by Kosovichev: see section 5 of Ref. 9). The neglect
of this temperature difference was a fundamental shortcoming of the
gas dynamic models of stellar \cite{Katsova} and solar
\cite{Livshits} flares in the opinion of the author
\cite{Mor16,Mor17}.

A possibility of attaining a concentration
$n_{\mathrm{H}}=3\cdot10^{16}\text{ cm}^{-3}$ (the value near the
photosphere \cite{Grin77}) owing to radiative cooling of the gas
behind the front of a plane-parallel stationary shock (propagating
``downward'' in the chromosphere of a red dwarf) was discussed in
\cite{Mor16} and the dissertation \cite{Mor17CSc} (assuming
\cite{Mor17CSc} no effect of the radiation field of the heated
layers lying near the photosphere). This is one of a ``set'' of
waves in the approach \cite{Bychkov19}. The author showed that: (a)
when the frozen-in condition for the magnetic field holds over the
entire radiative cooling time, an increase in the gas density by two
orders of magnitude is not possible; the corresponding rise in the
magnetic pressure $p_m$ by $10^4$ times halts (\cite{Mor17CSc}, pp.
90--91) the plasma compression; (b) when there is no coupling
between the changes in $n_{\mathrm{H}}$ and $p_m$, the increase in
$n_{\mathrm{H}}$ from $3.9\cdot10^{14}\text{ cm}^{-3}$ to
$3\cdot10^{16}\text{ cm}^{-3}$ \cite{Mor16} corresponding to the
hypothetical strong radiative cooling means that the gas flows out
of the viscous jump by a small distance $\Delta{}l_1\sim0.5{\text{
km}}$ \cite{Mor16} and for weaker cooling (to
$n_{\mathrm{H}}\approx10^{15}\text{ cm}^{-3}$), $\Delta{}l$
increases to $\Delta{}l_2\sim10{\text{ km}}$ \cite{Mor17CSc}
(results within the framework of a homogeneous plane layer model
\cite{Mor15}). Given the smallness \cite{Mor17CSc} $\Delta{l_1}$
compared to the linear sizes of the sources of the blue continuum of
the flares on AD Leo (dM4.5e) determined by Lovkaya \cite{Lov} in
the black-body approximation,\footnote{Based on the estimates of the
areas of the flares at the brightness maxima ($\sim10^{18}\text{
cm}^2)$ in a plane layer model.} in \cite{Mor16} it was stated that
the ``gas radiating behind a stationary shock front (propagating in
the direction of the photosphere of a red dwarf) is unable to
generate the [quasi-]black-body radiation … at the brightness
maximum of stellar flares.''\footnote{The radiative recombination
time $t_r$ of a dense chromospheric gas is low (\cite{Grin77}, p.
355) and the picture of radiative cooling behind the shock front is
independent of the choice of reference frame.} These results were
examined in \cite{Mor17CSc} as yet another argument in favor of the
viewpoint \cite{Grin77}.

The differences between the model of a stationary radiative shock
wave and the model \cite{Katsova} were noted in \cite{Mor16} and
\cite{Mor17CSc}. Thus, it was mentioned \cite{Mor16} that a system
of equations for one-dimensional gas dynamics in partial derivatives
was introduced in \cite{Katsova}, while the article of Belova {\it
et al.} \cite{Bychkov14} (used by the authors \cite{Bychkov19})
examines a system of ordinary differential equations (ODE) with
detailed accounting for elementary processes in the post-shock
plasma. Also in \cite{Mor17CSc} the neglect of thermal conductivity
($F_c=0$) responsible \cite{Syr1973} for the transfer of energy in
the high-temperature region of flares (initially heated by a beam of
non-thermal electrons to $T_e\sim10^7\text{ K}$, $T_i\sim10^6\text{
K}$ \cite{Somov81}) was noted (in \cite{Bychkov19} (p. 236), the ion
temperature $T_i$ of the gas is referred to as atom-ion one). In
addition, the increase in the geometric thickness $\Delta{z}$ of the
chromospheric condensation \cite{Katsova} (referred to as c.c. in
the following text) during impulsive heating was noted (a
non-stationary radiative shock wave \cite{Katsova} precedes a
temperature jump moving at a subsonic speed). Belova and Bychkov
\cite{Bychkov19} in criticizing the paper of Katsova {\it et al.}
\cite{Katsova} recall (see the section ``Discussion'') the first
result in \cite{Mor16}, and ignore the remarks\cite{Mor16,Mor17CSc}
on the differences in the models \cite{Katsova,Bychkov14}.

Finally, the origin of the $\mathrm{H}_{\alpha}$ line profile with a
blue asymmetry in the wings in the spectrum of a flare on UV Ceti
(dM5.6e) (Eason {\it et al.} \cite{Eason}) is discussed in Refs. 3,
8, and 13. Thus, it was noted that: (a) the model profile with a
Doppler core (half width $\Delta\lambda_D=0.9\text{ \AA}$
\cite{Eason}) and Stark wings ($\lg{n_e}=14.75$ \cite{Eason}) is
similar to the $\mathrm{H}_\alpha$ line profile (because the
parameter $b_{32}$ \cite{Mor15}\,$\ll1$); (b) the deviation from a
Stark profile for fitting of the right wing of the line (see Fig. 9
of Ref. 20) may be caused \cite{Sob95} by neglecting the
contribution of electron broadening to the formation of {\it
significantly opaque} (optically depth $\gg1$) wing of
$\mathrm{H}_\alpha$; (c) the Doppler core of $\mathrm{H}_\alpha$
\cite{Eason} is shifted as a unified whole to the left\footnote{It
is easy to confirm this by drawing two vertical segments from the
divisions corresponding to wavelengths of 6562 and 6564\,\AA{} in
Fig. 7а of Ref. 20 and by comparing the areas on the left and right
sides from the Doppler profile.} \cite{Mor16}. In \cite{Mor17} it is
stated that this kind of profile can be generated by the gas behind
a shock front which propagates ``upward'' in the partially ionized
chromosphere of a red dwarf (on the basis \cite{Mor16} that in a
laboratory reference system the core of a line in the plasma behind
a shock front should be ``shifted'' in the direction of motion of
the front).

The first part of this paper (I) contains a comparative analysis of
the approaches \cite{Katsova} and \cite{Bychkov19}. In section 2:
(a) it is argued that a ``set'' of stationary radiative shock waves
\cite{Bychkov19} cannot ensure simultaneous fulfillment of the heat
balance condition (between $P_e$ and the radiative energy losses
behind the shock front) and an increase in the thickness $\Delta{}z$
of the c.c. (cold gas ahead a thermal wave) --- the divergence
effect between the wave fronts --- during impulsive heating, as in
the model \cite{Katsova}; (b) it is demonstrated that the idea
(\cite{Bychkov17,Bychkov18}, \cite{Bychkov19}) of the formation of a
c.c. owing to radiative cooling of the plasma {\it in isolation
from} the thermal wave is not based on the fundamental article of
Kostyuk and Pikel'ner \cite{Pik74}, so that the critique of Belova
and Bychkov (\cite{Bychkov17,Bychkov18}, \cite{Bychkov19}) regarding
the paper of Katsova {\it et al.} \cite{Katsova} performed within
the framework of the approach of Ref. 22 (see Refs. 7 and 3) is
incorrect in the author's opinion; (c) the use \cite{Mor16} of a
fixed homogeneous plane layer of pure hydrogen plasma for an
approximate analysis of the radiative characteristics of the
high-density region \cite{Katsova} of thickness $\Delta{z_m}$
(instantaneous picture) is justified. In particular, it is
noteworthy that the claims \cite{Mor16} that the Menzel factors of
this layer are close to unity and of its transparency in the
continuum beyond the Balmer jump do not contradict Ref. 23 for gas
dynamic modeling by Allred {\it et al.} \cite{Allred} (the
populations of atomic levels are non-stationary explicitly).

In section 3 of this article it is shown that the ``set'' of shock
waves \cite{Bychkov19}, as opposed to models of the type of Ref. 7,
cannot heat the chromosphere of a red dwarf at significant
distances. In addition, it is argued that: (a) the calculations of
Belova and Bychkov \cite{Bychkov19} represent a development of the
kinematic model of solar flares (Nakagawa {\it et al.} \cite{Nak73})
and its application to dMe stars, specifically: a study of the
radiative response of the chromosphere to impulsive heating in the
simplest gas dynamic statement of the problem (a thermal wave is
excluded and a stationary approach is used); (b) as in \cite{Nak73},
the profiles of the ${\mathrm{H}}_\alpha$ line in the model
\cite{Bychkov19} conflict \cite{Pik74} with spectral observations;
(c) from the standpoint of the model \cite{Pik74}, the regions
behind the stationary shock fronts \cite{Bychkov19} correspond not
to the c.c., with time-varying thickness (in \cite{Katsova} from
$\sim1$ to $\sim10\text{ km}$), but to zones in which the plasma
relaxes to a state of thermal equilibrium; (d) the conclusion of
Refs. 8 and 15 that it is impossible to generate [quasi-]black-body
radiation (at the maximum brightness of stellar flares) behind the
front of one of these waves (velocity $u_0=60\text{ km/s}$
\cite{Mor16}) is confirmed by the calculations of Ref. 1. It is
emphasized that the separation (Belova and Bychkov \cite{Bychkov19})
of the model of Kostyuk and Pikel'ner \cite{Pik74} (basis of modern
gas dynamic program packages modeling the secondary processes in
solar and stellar flares) into ``thermal'' and ``shock-wave''
components is fundamentally impossible.

In the second part of this article (in preparation) it is argued
that both the blue and red components of the optical continuum of
stellar flares are formed near the photosphere \cite{Grin77} and the
point of view \cite{Bychkov19} according to which the source of
``hot'' quasi-black-body radiation is localized in the photosphere
of a red dwarf conflicts with observational data. Also: (a) the
effect of the radiation field of heated near photospheric layers (at
the maximum of the brightness of flares) on gas dynamic processes
taking place in the overlying layers of the chromosphere is
discussed in more detail than in \cite{Mor15,Mor16} and
\cite{Mor17CSc}; (b) the fundamental possibility is discussed of the
appearance and enhancement of $\mathrm{HeI}$ lines ({\it e.g.},
\cite{Eason}) in the thermal relaxation zone \cite{Katsova,Kos86}
(with increasing $T_e$ of the gas behind the front of a
non-stationary chromospheric shock wave owing to elastic collisions
of electrons with atoms and ions (($T_{ai}\gg{}T_e$)\footnote{For
high shock velocities corresponding \cite{Zharkova} to the high
fluxes $F_0$ in the non-thermal electron beams.}); and the origin of
the $\mathrm{H}_\alpha$ line profiles with a blue asymmetry in the
wings is discussed considering interpretations that differ from Ref.
13.\\
{\bf  2. Chromospheric condensation in the model of Ref. 7 and the
gas behind a stationary shock front.}

{\it{2.1.}} We begin by noting two fundamental differences in the
calculations \cite{Katsova} from the approach \cite{Bychkov19}:

(a) the system of gas dynamic equations \cite{Katsova} includes the
power of heating by a beam of accelerated electrons $P_e(\xi)$, the
classical thermal flux $F_c$ (Fourier law), gravitational
acceleration ($g={\mathrm{const}}$), and the cooling function $L(T)$
used over the entire range of the plasma temperature. In the
calculations \cite{Bychkov19} $P_e=0$, $F_c=0$, and $g=0$. Primary
attention is devoted to radiative cooling;

(b) the shock wave \cite{Katsova} is non-stationary (it propagates
in the chromosphere of a red dwarf in the direction of increasing
density, {\it i.e.}, ``downward''); in deep layers of the
chromosphere the wave transforms into a sound perturbation of a
discontinuous type (\cite{Pik74}, p. 594). In the approach of Ref. 1
 the shock velocity $u_0=\mathrm{const}$ and the pre-shock gas is
uniform (a so-called stationary shock wave). The problem of
accounting for the density gradient in the chromosphere is solved
palliatively in Ref. 1; they consider a range of values of $u_0$
from $30$ to $100\text{ km/s}$ and obtain a corresponding ``set'' of
profiles of stationary radiative shock waves.

 The difference ``a'' shows up in that in the model of Ref. 7
the thermal wave is responsible for production of the compression
wave ahead of itself, and after some time (at the heating of the
denser plasma) it becomes a shock; ``the thermal front then acts as
a piston pushing the gas'' \cite{Kos86} (a so-called temperature
wave of the second kind \cite{Vol63}). As a result of the emission
from the post-shock gas of the non-stationary shock wave, a layer of
dense cold plasma appears ahead the temperature jump with a typical
flat profile of $T$ (Fig. 1): the radiative cooling ceases when the
loss of energy owing to radiation $L(T)$  becomes comparable to the
energy influx $P_e$ from the thermal wave (the transition of the
shock wave in the {\it stationary} regime \cite{Kos86}).

During impulsive heating, the fronts of the thermal and shock waves
diverge (the temperature jump moves at a subsonic speed
\cite{Katsova,Vol63}), so that the geometric thickness $\Delta{z}$
of the cold gas (after passage of the shock wave), the c.c.,
increases. (It is clear from Fig. 1 how the region of elevated
density is established at an ever higher range of values of $\xi$.)
As a result, closer to the end of heating, the width of the thermal
relaxation zone $\Delta{l_1}\sim0.5\text{ km}$
\cite{Mor16}\footnote{An increase in $n_{\mathrm{H}}$ by two orders
of magnitude behind the front of the downward moving (toward the
photosphere) shock is discussed in Ref. 7. At the same time, an
increase in the concentration by a factor of $\approx77\sim100$ is
``hidden away'' in the estimate of $\Delta{}l_1$ \cite{Mor16}.}
turns out to be small compared to $\Delta{}z_m$.

The authors \cite{Bychkov19} assume that the ``set'' of stationary
radiative shocks in the chromosphere of a red dwarf exists
independently of the thermal wave ($P_e=0$). As a result, the
regions behind the fronts of the shock waves \cite{Bychkov19} in the
calculations of Ref. 7 correspond formally to the zones where the
plasma relaxes to a state of thermal equilibrium, with a sharp
increase in $n_\mathrm{H}$ owing to radiative cooling (see Fig. 1).

The difference ``b'' shows up in that in the approach of Ref. 1, the
accounting for the radiative losses of the gas behind a shock front
is substantially more precise than in the model of Ref. 7. This
result is attained through substantial simplifications in the gas
dynamic part of the problem statement \cite{Bychkov19} compared to
that of Ref. 7: the thermal wave is excluded and a {\it stationary}
approach is used (see section 10 in \cite{Kaplan}). In addition, in
\cite{Bychkov19}, $T_{ai}$ and $T_e$, the concentrations of the
plasma components, and the relative populations of the $\mathrm{HI}$
levels ($\nu_k=n_k/n_{\mathrm{H}}$) are found as functions of $t$ by
solving a system of ODE for $\cfrac{d\nu_k}{dt}\neq0$, the
ionization states of the gas components, the internal energy of the
plasma, and the derivative ${dT_e}/{dt}$ (Eqs. (32), (40), (41),
(57), and (89) in \cite{Bychkov14}) and a number of auxiliary
algebraic equations.\footnote{The calculations of {\it Ref. 17} used
  \cite{Mor17} a two-level approximation: it assumes that $\nu_2(t)=4\nu_1(t)[1+\Omega_{21}^{-1}]^{-1}\cdot\\
  \cdot\exp(-\Delta{}E_{12}/k_{\mathrm{b}}T_e)$, where $\Omega_{21}=q_{21}n_e/{A_{21}^*}$. Here
  $\Delta{E_{12}}$ is the excitation energy of the second level
  relative to the ground one,
  $k_{\mathrm{b}}$ is the Boltzmann  constant, $q_{21}$ is the electron impact de-excitation coefficient for the hydrogen atom, and
  $A_{21}^*$ is the effective spontaneous transition probability.} Here $t$ is the time since
  the time a given element of the gas intersected the shock front. The shock is stationary, so
$d{\widetilde{l}}=u(t)dt$, where ${\widetilde{l}}$ is the distance
from the viscous jump, and $u(t)$ is the velocity of the gas in the
reference system associated to the discontinuity at time $t$. (A
detailed derivation of the system of equations for calculating the
profile of a stationary radiative shock (under the conditions of the
partially ionized chromospheres of dMe stars) including the effect
of the radiation field of the heated near photospheric layers is
given in Ref. 15, pp. 22--28, 71--90.)

Given these remarks, we have the following differences in the
non-stationary cooling of the plasma behind the shock fronts in the
models discussed here:

(a) Ref. 7: the radiative cooling is caused by the inequality of the
heating and cooling functions owing to the sharp rise in $T$ and
$n_{\mathrm{H}}$ of the plasma; a reduction in the gas temperature
and a rise in its density owing to emission continue until a thermal
balance is established between the influx of energy from the
non-thermal electrons and loss owing to radiative cooling;

(b) Ref. 1: the plasma parameters corresponding to the cold gas are
specified by the {\it choice} of the final step in $t$. The increase
in plasma density is limited only by the pressure rise of the
magnetic field and the influx of energy from ``photospheric
emission'' \cite{Bychkov19} with a temperature $<6\cdot10^3{\text{
K}}$ \cite{Bychkov19}.

Thus, the ``set'' of shock waves \cite{Bychkov19} cannot ensure
simultaneous satisfaction of the condition for heat balance and an
increase in the thickness of the c.c., as  occurs in Ref. 7.

   {\it 2.2.} Belova and Bychkov \cite{Bychkov19} expound their concept of the calculations of the type
   \cite{Katsova}
   (see their critique of the work of Allred {\it et al.} \cite{Allred}, in Refs. 6 and 1) in the
following way: ``in \cite{Katsova}, a {\it hypothesis} of
``chromospheric condensation'' is formulated, according to which
black-body radiation originates from a region of size about
$10\text{ km}$ located at a height of roughly $15000\text{ km}$ (so
in \cite{Bychkov19}) and formed by gas that has been isobarically
compressed by radiative cooling behind a shock front to a
temperature of about $9000\text{ K}$ and density of about
$10^{15}\text{ cm}^{-3}$.'' Likewise, in justification of their
views, the authors \cite{Bychkov19} introduce a citation from the
abstract of the pioneering article of Kostyuk and Pikel’ner
\cite{Pik74}: ``a temperature jump that is primarily associated with
thermal conductivity propagates through the gas. A shock wave
propagates ahead of the temperature jump, the wave heats and
compresses the gas. The velocity of motion decreases with depth.''

Here Belova and Bychkov \cite{Bychkov19} omit the following sentence
from the abstract for that article: ``in this paper, the viscosity
is neglected, and heating by the shock does not appear explicitly,
but it is estimated from the Hugoniot adiabatic curve.'' Thus, ``the
lower front of the motion'' in the {\it calculations} of Ref. 22
``is not a shock wave...'' (\cite{Pik74}, p. 594). Therefore, in
\cite{Pik74}, the divergence of the fronts of the thermal and {\it
shock} waves during a time of impulsive heating could not be
described, as it is in Ref. 7, where an artificial (quadratic)
viscosity $\omega$ \cite{Katsova,Kos79} is introduced; the authors
\cite{Pik74} hoped to ``solve the problem with viscosity, at a
velocity downward to 100 km/s,'' later \cite{Pik74}. This conclusion
is confirmed on comparing the geometric thickness $\Delta{}z$ of the
region in which the $\mathrm{H}_\alpha$  line emission (the
conditions in the solar chromosphere) is predominantly localized; in
Ref. 22 it is 8 km \cite{Kostyuk} (pulse duration 100 s, ``an
illusion of continuity'' caused by poor time resolution
\cite{Somov79}), while in Ref. 14 it goes to $\sim10{\text{ km}}$
over $10{\text{ s}}$ of impulsive heating (a single elementary flare
burst, EFB).

The authors \cite{Bychkov19} also do not consider the following
comment of Kostyuk and Pikel’ner \cite{Pik74} (p. 593) regarding the
{\it qualitative} picture of processes in the region of the optical
burst (in $\mathrm{H}_\alpha$): ``the gas heated by the shock cools
{\it sooner} than the temperature jump reaches it,'' so that ``the
temperature has a minimum between two heated regions.'' Also
\cite{Pik74}: (a) the shock speed (``downward'') depends on the
energy flux $F_0$ in the beam of accelerated electrons, the
radiative energy loss per $\text{ cm}^2$ of the region ahead the
thermal wave (the {\it cold} radiating gas), the density of the
unperturbed chromosphere, and the plasma temperature in the hot
region (at the jump); (b) {\it part} of the c.c. is similar
\cite{Pik74} to the flare element in the model of Brown
\cite{Brown}, where the energy influx from the non-thermal electrons
is balanced by
 radiative losses in $\mathrm{H}_\alpha$.

In addition, the simplified representation \cite{Bychkov19} about a
temperature jump ahead of which a non-stationary radiative shock
propagates does not fully reflect the complicated interaction of the
thermal and shock waves in the calculations of Ref. 7. Thus, it is
noted \cite{Katsova} that ``over a certain time the thermal wave
‘amplifies’ the shock wave.''

Therefore, the claim by Belova and Bychkov
\cite{Bychkov18,Bychkov17,Bychkov19} that the formation of the c.c.
is caused exclusively by radiative cooling (``in a separation'' from
the thermal wave) is not based on the Kostyuk-Pikel’ner model
\cite{Pik74}. (We recall that in Ref. 1 $P_e=0$, $F_c=0$, $g=0$, and
$Q=0$.) For this reason, the critique
\cite{Bychkov18,Bychkov17,Bychkov19} regarding the paper of Ref. 7
in terms of this model  is not correct.

{\it 2.3.} Referring to the book \cite{iva69}, Katsova {\it et al.}
\cite{Katsova} assume that in sufficiently dense layers of the
unperturbed chromosphere, where the optical depth in the resonance
transition $\geq{}\tau_{cr}=10^6$ \cite{Katsova}, the populations of
hydrogen levels are determined by the Boltzmann formula
\begin{equation}\label{Eq1}
\cfrac{n_2}{n_1}=4\cdot\exp\left(-\cfrac{\Delta{}E_{12}}{T_{\mathrm{eV}}}\right)\equiv{}f(T),
\end{equation}
where $T_{\mathrm{eV}}$ is the gas temperature in electron-volts.
Here \cite{Katsova} the degree of ionization of the plasma
$x^*={n_e}/{n_{\mathrm{H}}}$ \cite{Katsova} and $n_2$ are found by
solving the system of equations (\ref{Eq1}) and (\ref{Eq2}):
\begin{equation}\label{Eq2}
n_1n_eq_1+n_2n_eq_2=n_e^2\widetilde{\alpha},
\end{equation}
where $q_1$ and $q_2$ are the electron-impact ionization
coefficients of the atom from levels 1 and 2, and
$\widetilde{\alpha}$ is the total coefficient of spontaneous
photorecombination to all levels except the first (an approximate
accounting for $L_c$ radiation scattering). $n_\mathrm{H}$ and $T$
are assumed to be specified.\footnote{It should be noted that Eqs.
(\ref{Eq1}) and (\ref{Eq2}) contradict one another: the Saha
equation should be used (as in Ref. 22) instead of Eq. (\ref{Eq2}),
neglecting the effect of the radiation field of the photosphere of a
red dwarf.}

In Ref. 7 it is stated that the heating and cooling functions ($P_e$
and $L$), classical thermal flux $F_c$, pressure $p$, and energy
$\varepsilon$ were calculated ``simultaneously with the degree of
ionization of the hydrogen plasma''.\footnote{In the model
\cite{Katsova}, in the upper layers of the chromosphere the quantity
$x^*$ was ``replaced by the analogous quantity'' $x$, defined within
the framework of the ``one level + continuum'' model of an atom.}
For this reason, on one hand the relation of $n_e$ and
$n_{\mathrm{H}}$ in the c.c. should be determined from Eqs.
(\ref{Eq1}) and (\ref{Eq2}), and on the other, a balance
\cite{liv83} between the energy influx from the thermal wave
($\xi>\xi_0=E_{10}^2/2a$, where $a$ is a function of $E_{10}$ and
$n_{\mathrm{H}}$ \cite{Syr72}) and the radiative losses in the
$\mathrm{H}_\alpha$ line ($L_{{\mathrm{H}}_\alpha}$) must be
ensured: a ``manifestation'' of the flare element of Brown
\cite{Brown} in the model \cite{Pik74}.

Following Ref. 33 we check fulfillment of these statements in Ref.
7. We begin with the fact that $0.8\text{ s}$ have passed from the
beginning of impulsive heating (the fronts of the thermal and shock
waves have not separated too strongly); we do not examine
\cite{liv83} the plasma layers on reaching the temperature jump
since in this part of the c.c. there should be movements such as an
“upward” expansion \cite{Brown}.

Solving the system of Eqs. (\ref{Eq1}) and (\ref{Eq2})  yields
$x^*=\left\{1+\widetilde{\alpha}[1+f(T)]/[q_1+q_2f(T)]\right\}^{-1}$;
it is clear that $x^*=x^*(T)$. $T\approx9000{\text{ K}}$
\cite{Katsova} ($T_{\mathrm{eV}}\approx0.78{\text{ eV}}$), so that
$f(T)\approx7.7\cdot10^{-6}\ll1$ ({\it i.e.}, $n_2\ll{}n_1$) and
$x^*\approx0.03$. Here as in Refs. 7 and 14, the coefficient
$\widetilde{\alpha}$ was calculated according to Seaton, but $q_1$
and $q_2$, according to Johnson \cite{Johnson}
($q_1\ll{}q_2$).\footnote{In general, in the range of $T$ from
$10^4{\text{ K}}$ to $2.5\cdot10^4{\text{ K}}$  of astrophysical
interest, $q_k$ {\it increases} \cite{Shn} as the principal quantum
number gets larger.} From Eqs. (\ref{Eq1}) and (\ref{Eq2}) we also
find that $n_2\approx(1-x^*)n_{\mathrm{H}}f(T)$. A Lagrange variable
of $\xi_1=10^{20}\text{ cm}^{-2}$ corresponds to
$n_{\mathrm{H}}\approx6\cdot10^{15}\text{ cm}^{-3}$ (see Fig. 1),
whence $n_2\approx4.5\cdot10^{10}\text{ cm}^{-3}$.

In turn \cite{Katsova},
\begin{equation}\label{Eq3}
P_e(\xi)/n_{\mathrm{H}}=0.7x^*P_1(\xi)+0.3P_1(\xi),
\end{equation}
where $P_1(\xi)$ is the Syrovatskii and Shmeleva function
\cite{Syr72} (the so-called CEA approach \cite{Zharkova}); for
$\gamma=3$, $E_{20}\rightarrow\infty$ ($E_{20}$ is the upper
boundary of the accelerated electron spectrum), $\xi{}\geq{}\xi_0$
$P_1(\xi)=2^{-3.5}\pi{}F_0{}E_{10}a^{-0.5}\xi^{-1.5}$ \cite{Syr72}.
The value of $P_1(\xi_1)=1.9\cdot10^{-9}\text{ erg/s}$. From Eq.
(\ref{Eq3}) we finally obtain
$P_e(\xi_1)\approx6.1\cdot10^{-10}n_{\mathrm{H}}$.

On the other hand, the radiative energy loss in the
$\mathrm{H}_\alpha$ line \cite{Katsova} is:
$L_{\mathrm{H}_\alpha}\approx{}0.14n_2{}x^*q_{23}n_{\mathrm{H}}\mathrm{Ry}$;
taking $q_{23}\approx5\cdot10^{-8}\text{  cm}^3{\text{/s}}$ (the
electron-impact excitation coefficient for the atom) \cite{Johnson}
, we have
$L_{\mathrm{H_\alpha}}\approx1.9\cdot10^{-10}n_{\mathrm{H}}$. (We
neglect the contribution of $L_{{\mathrm{L}}_\alpha}$ \cite{liv83}
since the escape probability for a resonance photon outside of the
c.c. is low.)

The absence of a divergence between $P_e(\xi_1)$ and
$L_{\mathrm{H}_\alpha}$ by orders of magnitude serves as a
quantitative confirmation that the calculations of Ref. 7, as
opposed to those of Belova and Bychkov \cite{Bychkov19} which
exclude a thermal wave, have been done \cite{Katsova} within the
framework of the model of Kostyuk and Pikel’ner \cite{Pik74}.

{\it 2.4.} The authors of Ref. 7 analyze the characteristics of the
emission from the c.c., with time-varying thickness (in particular,
for $\Delta{}z_m$), using the results of Grinin and Sobolev
\cite{Grin77} for a fixed homogeneous plane layer of pure hydrogen
plasma with a source function $S_\nu=B_\nu(T)$. Thus, Katsova {\it
et al.} \cite{Katsova} set up a correspondence between the
“densifications” \cite{Katsova} and the “set” of plane layers
(instantaneous pictures)\footnote{With this approach, information on
the distribution of the velocities, “both upward and downward”
(\cite{Pik74}, p. 596), in the c.c. is lost; the motion of the
layers set up in Ref. 7 in accordance with “downward”
“densifications” is ignored \cite{liv83}.}, while simultaneously
{\it correcting} $x^*$ by the equilibrium value $x^{\mathrm{eq}}$
for {\it fixed} $n_{\mathrm{H}}$, $\Delta{}z$, and $T$ (the
populations of the atomic levels of the gas in the c.c. correspond
to the Boltzmann formula in view of Eq. (1) and the above discussed
features of the problem statement in Ref. 7; for $T\approx9000\text{
K}$, $n_{\mathrm{H}}=2\cdot10^{15}\text{ cm}^{-3}$, and
$\Delta{z_m}$ $x^{\mathrm{eq}}\approx0.15$).

Numerical solution of the system of balance equations for the
elementary processes \cite{Mor15} shows \cite{Mor16} that within a
motionless homogeneous plane layer of thickness $\Delta{z_m}$  with
$T_{ai}=T_e$  (in Ref. 7 the cooled and, thus, single-temperature
\cite{Mor16} gas behind the thermal relaxation zone) the Menzel
factors of the plasma differ little from unity ({\it i.e.},
$S_\nu\approx{}B_\nu(T)$): given the high density of the gas, its
low temperature, and high optical depth in spectral lines (for the
resonance transition at the center of the layer
$\tau_{\mathrm{L}_\alpha}\approx3\cdot10^7$  –- see Eq. (53) in Ref.
3) the radiative processes turn out to be secondary \cite{Mor15}
compared to collisional ones. In particular, for {\it conservative}
scattering the average escape probability over the layer for a
resonance photon beyond the confines of the plasma (in the case of a
symmetric model profile of the spectral line with a Doppler core and
Holtsmark wings) is $\theta_{\mathrm{L}_\alpha}\sim10^{-6}\ll1$
\cite{Mor15,Mor16} (recall that $\theta_{\mathrm{L}_\alpha}$
obtained with an approximate analytic solution \cite{iva72} of the
radiative transfer equation, for $\tau_{\mathrm{L}_\alpha}\gg1$
turns out to be close \cite{iva09} to the escape probability for a
photon beyond the confines of the plasma without scattering from the
center of the layer). The {\it smallness} of
$\theta_{\mathrm{L}_\alpha}$ also implies the secondary importance
of including the motion of the layer as a unified whole in the
direction toward the photosphere when solving the system of steady
state equations \cite{Mor15} (but without calculating the
$\mathrm{H}_\alpha$ line profile \cite{Pik74,Kostyuk}).

{\it  2.5.} In the article of Allred {\it et al.} \cite{Allred}, the
response of the chromosphere of the Sun and dMe stars to impulsive
heating by a beam of non-thermal electrons has been modeled by
combined solution of the equations of radiative gas dynamics
(plane-parallel approximation; $g\neq0$), the equations for {\it
non-stationary} \cite{Canf83} atomic level populations, and the
radiative transfer equation. For the questions discussed here the
important fact (\cite{Kow18}, pp. 9, 10) is that for fluxes $F_0$
equal to $5\cdot10^{11}\text{ erg/cm}^2{\text{s}}$ (the Sun’s
chromosphere) and $10^{13}\text{ erg/cm}^2{\text{s}}$ (dMe stars;
referred to below as model $F13$), the populations of hydrogen
levels in chromospheric condensations are close to the LTE values
{\it after some time} (see table 2 in Ref. 23) after the onset of
impulsive heating, except for the upper part (a thickness of
$\sim1{\text{ km}}$) for each condensation (zone behind the
non-stationary shock front), where the populations of levels with
$k=1,2$ {\it deviate significantly} from their equilibrium
values.\footnote{We note that the definition of c.c. given in Ref. 9
differs from that used in Ref. 23.}

From a physical standpoint these results are caused by the
following:

(a) including in the conservation law of the internal energy of the
gas \cite{Allred}: the divergence of the radiative flux
${\partial{}F_r}/{\partial{}z}$ ($z$ is the height per unit mass)
and the power of the plasma heating by the beam of accelerated
electrons (enters with the {\it opposite} sign) which ensure
fulfillment of the condition for thermal balance, and the divergence
of the thermal conductivity flux (same sign as
${\partial{}F_r}/{\partial{}z}$);

(b) the high density \cite{Kow18} of the gas in the chromospheric
condensations, their substantial geometric thickness ($\sim$ several
tens of km along the vertical \cite{Kow18}) and the comparatively
low $T$ of the plasma ahead the thermal wave (in the $F13$ model
$T\sim10^4{\text{ K}}$ \cite{Kow18}).

The motionless homogeneous plane layer of thickness $\Delta{}z_m$
with $n_{\mathrm{H}}=2\cdot10^{15}\text{ cm}^{-3}$,
$T\approx9300\text{ K}$ produces radiation with an optical depth at
a wavelength of $4170$ \AA{} $\tau_{4170}\approx0.02$
\cite{Mor15,Mor16}. On the other hand, in Ref. 23, $\approx2\text{
s}$ after the onset of heating (calculation with $F_0=10^{13}\text{
erg/cm}^2\text{s}$, $\gamma=3$, $E_{10}=37\text{ keV}$)
$\tau_{4170}\approx0.5$ in [the lower part of] a c.c. of thickness
$\sim20\text{ km}$ с $n_{e_{\mathrm{max}}}\approx5\cdot10^{15}
\text{ cm}^{-3}$ at $T\sim10^4{\text{ К}}$.

Thus, the conclusion of Ref. 8 regarding the transparency of the gas
($\tau_{4170}\ll1$) in a layer with parameters corresponding to the
c.c. \cite{Katsova} (model $F12$) {\it does not conflict} with the
results of Ref. 23 obtained by gas dynamic modeling (the populations
of the atomic levels are non-stationary explicitly) and is, thereby,
positive.\footnote{The inconsistency of Ref. 14 regarding the nature
of white flares on dMe stars with the results of RADYN in the
section on the value of the energy flux $F_0$ was first pointed out
by Kowalski \cite{Kow12}.}

{\bf 3. Shock-wave model of stellar flares.} The characteristic time
for recombination radiation of an optically {\it thin} gas behind a
stationary shock front (one of the “set” of waves in the approach of
Ref. 1) is
\begin{equation}\label{Eq4}
t_r\sim(\alpha\cdot{}n_e)^{-1},
\end{equation}
where $\alpha$ is the spontaneous radiative recombination
coefficient summed over the levels $k=2,…,k_{\mathrm{max}}$
($k_{\mathrm{max}}$ is given by the Inglis-Teller criterion).
Thence, for $n_e=10^{14}\text{ cm}^{-3}$, $T_e=10^4\text{ K}$ and
using the Kramers approximation, we obtain
$t_r\approx4\cdot10^{-2}\text{ s}$.

On the other hand, the time for evolution of the c.c. \cite{Katsova}
(the cold gas ahead a thermal wave) is $t_h\approx9\text{ s}$ (see
Fig. 1), comparable to the duration of all the impulsive heating,
$10\text{ s}$. Furthermore, after heating ceases ($t>10\text{ s}$)
the shock \cite{Katsova} is rapidly damped \cite{Kos86}. Thus, the
time of existence and penetration depth into the chromosphere of the
non-stationary shock are first of all determined by the presence of
an influx of energy from accelerated electrons (through the thermal
wave), {\it i.e.}, the duration of the impulsive heating
\cite{Somov79,Katsova}. Remembering that the duration of the line
emission in flares ranges from several minutes and longer
\cite{Gersh77} and given that a flare consists of a “set” of
elementary events (for example, Ref. 20), we find that
$t_h\sim10\text{ s}$ is a good agreement between theory and
observations. Thus, without a constant influx of energy from
outside, the “set” of shock waves \cite{Bychkov19} cannot exist for
a long time.

Finally, multiplying $t_r$ by $u_0=100\text{ km/s}$, we find that
over the radiative cooling time, the corresponding shock wave
\cite{Bychkov19} travels a path of $\approx4\text{ km}$. Therefore,
the “set” of shock waves \cite{Bychkov19} cannot heat the
chromosphere of a red dwarf over significant distances.

For these reasons it is impossible to agree with the claim of Belova
and Bychkov \cite{Bychkov19} that the “model of a shock wave
propagating in the gas of the chromosphere can be used to explain
... emission of the hydrogen plasma that is transparent at continuum
frequencies.” The component of the emission from stellar flares that
is transparent in the continuum is {\it mainly} formed in the c.c.,
but {\it not only} \cite{Bychkov19} in the zone behind the front of
the non-stationary shock wave where the plasma relaxes to a state of
thermal equilibrium.

We note that the above comments against the approach of Ref. 1 are
known in the scientific literature \cite{Somov79,Somov80,Fisher85}
as applied to the kinematic model of solar flares (Nakagawa {\it et
al.} \cite{Nak73}), where it is proposed that a “set” of stationary
shock waves that heat and compress the gas  ($g\neq0$,
single-temperature post-shock heating (with $T_{ai}=T_e$), the
cooling function with a power law dependence on $T$) and, thereby,
intensify the $\mathrm{H}_\alpha$ radiation \cite{Pik74}, propagates
in the chromosphere. In fact, the calculations of Ref. 1 represent a
development of the model of Ref. 25 and its application to dMe
stars, specifically: a study of the radiative response of the
chromosphere of a red dwarf to impulsive heating in the simplest gas
dynamic formulation of the problem (a thermal wave is excluded and a
{\it stationary} approach \cite{Kaplan} is used).

An inconsistency of the $\mathrm{H}_\alpha$ line profile in the
model of Ref. 25 with observational data was noted in the papers by
Canfield and Athay \cite{CanfAth74} and Kostyuk and Pikel’ner
\cite{Pik74}: according to the calculations of Ref. 43, the profile
“has a strong, shifted central reversal, so that in Ref. 43 it was
necessary to assume turbulent motions at velocities of 40--70 km/s”
\cite{Pik74}. From a physical standpoint these defects of the
profile are caused, on one hand by the large optical depth in the
center of the $\mathrm{H}_\alpha$ line core and, on the other, by
the significant velocity of the gas behind the stationary shock
front in the laboratory coordinate system: $u_1<0.75u_0$ (we neglect
radiative cooling). Therefore, the $\mathrm{H}_\alpha$ line profiles
in the approach of Ref. 1 {\it contradict} the observational
data.\footnote{The optical depth in the resonance transition in the
radiative cooling region of the gas behind a shock front
\cite{Bychkov19} propagating at a velocity of $u_0=60\text{ km/s}$,
$\tau_{\mathrm{L}_\alpha}\sim10^7$ \cite{Mor15} (the value of
$\tau_{\mathrm{L}_\alpha}$ is reckoned from the viscous jump).}

It was also noted \cite{Pik74} (p. 598) that in the approach
\cite{Pik74} the $\mathrm{H}_\alpha$ profile is a “combination”
\cite{Pik74} of the profile corresponding to the gas ahead a heat
wave (the Brown model \cite{Brown}: a symmetric profile with a deep
reversal in the center \cite{Canf74}) and a line profile in the
kinematic model \cite{Nak73}. Since the “maximum of one profile is
superimposed to the minimum of the other ... a strong reversal is
not obtained” \cite{Pik74}. In other words, the $\mathrm{H}_\alpha$
line will have a symmetric core with a shallow “dip” in the center
and a red asymmetry of the wings \cite{Ich}.

The above remarks imply that the model of Kostyuk and Pikel’ner
\cite{Pik74} lying at the basis of modern gas dynamic program
packages for modeling secondary processes in solar and stellar
flares does not {\it in principle} allow “splitting” (Belova and
Bychkov \cite{Bychkov19}) into “thermal” and “shock wave”
components; otherwise it “degenerates” into the model of Nakagawa
{\it et al.} \cite{Nak73}. Therein lies the main difficulty.

Nevertheless, calculating the profiles of stationary radiative
shocks is of interest \cite{Mor17CSc} from the standpoint of support
of views \cite{Grin77} regarding the nature of the blue component of
the optical continuum at the brightness maximum of powerful flares
on dMe stars, since it is precisely in the thermal relaxation zone
the plasma density $n_{\mathrm{H}}$ increases sharply owing to
radiative cooling. The conclusion \cite{Mor16} that the
quasi-black-body radiation cannot be generated in flares by
post-shock gas of one of a “set” of shock waves \cite{Bychkov19}
(velocity $u_0=60\text{ km/s}$) under the conditions of the
chromospheres of red dwarf stars is based on the optical depth in
the $\mathrm{L}_\alpha$ line ($\sim10^7$ \cite{Mor15}) and the
hydrogen atom concentration in the ground state
($2\cdot10^{16}\text{ cm}^{-3}$, the hypothetical strong radiative
cooling regime) in the region where the gas radiates. This
conclusion does not use the approximation of stationary populations
behind the shock front, so it is correct. It is now confirmed
\cite{Bychkov19} by direct calculation.\footnote{It is necessary,
however, to caution against attempts to treat Ref. 1 as a
calculation of an improved cooling function that can be used in the
complete system of gas dynamic equations \cite{Katsova} instead of
the corresponding function $L(T)$: the theory of stationary
radiative shock waves is an {\it independent} area of radiative gas
dynamics.}

{\bf 4. Additional comments.} We emphasize that, as opposed to Refs.
11 and 8, in Refs. 46 and 23, interpretations are given not only of
the blue but also of the NUV components of the continuum spectrum
during the impulsive phase of stellar flares, as well as of the
energy distribution over wavelengths of $3646-3730\text{ \AA}$. For
this, Kowalski \cite{KowIAU} uses a “composite” model of a perturbed
chromosphere of a red dwarf that includes a c.c. ({\it flux F13})
and stationary layers with $T\sim9000-12000\text{ K}$ and a
thickness of a few hundred km lying ahead of the front of a
non-stationary shock. These layers are heated \cite{Kow18} by high
energy ($E_e\gg{}E_{10}$) electrons from a beam with a falling
power-law spectrum –- a mechanism pointed out in Ref. 22 (p. 594).

The difficulties with the F13 approach owing to the so-called
reverse current problem ({\it e.g.}, Ref. 19) are noted in Ref. 23.
In addition, the number of electrons with $E_e\gg{}E_{10}=37\text{
keV}$ \cite{KowIAU} is {\it substantially lower} than with a kinetic
energy near $E_{10}$.

The authors of Ref. 23 believe that stationary layers can make a
significant contribution to the continuum spectrum of flares if the
optical thickness in the c.c. is $\ll1$. It is clear that this point
of view differs from that discussed in Refs. 11 and 8. At the same
time, a comparison of the theoretical (for different radiation
mechanisms) and observed color indices implies \cite{Zhil} that the
best agreement is obtained for a combination of “short-lived
[quasi-]black-body radiation near the maxima of a flare and the {\it
long-lived} radiation from a hydrogen plasma with temperatures and
densities {\it somewhat higher} than in the unperturbed
chromosphere.”

Belova and Bychkov \cite{Bychkov19}  assume that their calculations
“... do not confirm the {\it hypothesis} advanced in Ref. 7 of a
bright optically thick, in the continuum, “chromospheric
condensation,” formed during a flare by radiative cooling of the
post-shock gas.” As noted above, the regions behind the fronts of
the stationary shocks \cite{Bychkov19} in the model of Ref. 7
correspond formally to zones in which the plasma relaxes to a state
of thermal equilibrium (see Fig. 1) and not to a dense cold gas
ahead a thermal wave (a c.c. of thickness $\Delta{}z_m$). For this
reason, the calculations \cite{Bychkov19} cannot serve as a
replacement for the analysis \cite{Mor16} of the characteristics of
the continuous optical radiation from a layer with parameters
corresponding to a “densification” \cite{Katsova}, just as they
cannot supplement the selection \cite{Mor17CSc,Mor17} of errors by
Katsova \cite{Kats81} in Refs. 7 and 14.

{\bf 5. Discussion.} The  analysis in sections 2 and 3 of this
article makes it possible to discuss in more detail than in
\cite{Mor16} and \cite{Mor17}, the statement of the problem in
\cite{Mor15} and the results obtained there. The authors of Ref. 3
calculated the spectrum of the radiation from a homogeneous plane
layer of pure hydrogen plasma with $T_{ai}\neq{}T_e$ passing through
the front of a stationary shock wave ({one of the ``set'' of waves
in the approach of Ref. 1). The layer was assumed to be motionless,
since the post-shock gas moves at a {\it subsonic} speed relative to
the viscous jump; no precursor was included. The physical parameters
of the layer ($T_{ai}$, $T_e$, and $\tau_{\mathrm{L}_\alpha}$) were
chosen on the basis of a calculation \cite{Bychkov14} of the profile
of a stationary radiative shock under the conditions of the
atmospheres of the variable stars of the type {\texttt{o Cet}} (the
density of the gas ahead of the front is $n_0=10^{12}\text{
cm}^{-3}$, $u_0=60\text{ km/s}$). A range of $n_{\mathrm{H}}$ from
$3\cdot10^{14}\text{ cm}^{-3}$ to $3\cdot10^{16}\text{ cm}^{-3}$ was
examined.

Reference 3 is a study of the influence of the post-shock inequality
\cite{Pik54} for the atom-ion and electron temperatures of the
plasma on the formation of the emission spectrum of this kind of
layer. We note that from the standpoint of the theory of stationary
radiative shock waves these calculations are abstract: this is
indicated by the {\it invariance} of the values of the atom-ion and
electron temperatures in the confines of the layer \cite{Mor15} and
by the solution of the {\it balance} equations for the elementary
processes involving the atomic level populations.

The author \cite{Mor16} has confirmed that the results of Ref. 3 are
correct even in the {\it single-temperature} case, {\it i.e.}, when
$T_{ai}=T_e$.\footnote{The fact that the homogeneous layer
\cite{Mor15} generates a continuous spectrum that is close to a
black-body spectrum (see Fig. 2 of Ref. 3) indicates that the
difference between $T_{ai}$ and $T_e$ \cite{Pik54,Mor15} had no
significant influence on the formation of the emission spectrum of
this kind of layer.} In particular, the following remain unchanged:
(а) Eq. (71) for the full intensity of the radiation in the Balmer
series lines including the fact that for certain shifts in frequency
from the center, the wings of the lines are ``immersed'' (Eason {\it
et al.} \cite{Eason}, p. 1167) in the continuum (here the asymmetry
of the wings was neglected for simplicity); (b) the criterion for
applicability in the calculations of a symmetric model profile with
a Doppler core and Holtsmark wings (Eq. (49) in Ref. 3}).

We emphasis that the c.c. in the model \cite{Katsova} -- a {\it
single-temperature} layer (cold radiating gas ahead a thermal wave)
-- and a two-temperature layer \cite{Mor15} are ``unified'' only by
the formal closeness of $\Delta{}z_m$ to $\Delta{}l_2$ or to the
layer thickness \cite{Mor15} of $10\text{ km}$.

{\footnotesize Further, a number of corrections and refinements to
bring the article \cite{Mor16} into line with \cite{Mor17} are
given.}

{\small The author thanks Drs. Yu. A. Fadeyev and N. N. Chugai for
useful comments made during discussions of the basic results of
Refs. 3 and 8 and the dissertation \cite{Mor17CSc} during the
Astrophysics Seminar at the Institute of Astronomy of the Russian
Academy of Sciences (INASAN).

\newpage

\end{document}